\documentclass[lettersize,journal]{IEEEtran}
\usepackage{amsmath,amsfonts}
\usepackage{algorithmic}
\usepackage{algorithm}
\usepackage{array}
\usepackage[caption=false,font=normalsize,labelfont=sf,textfont=sf]{subfig}
\usepackage{textcomp}
\usepackage{stfloats}
\usepackage{url}
\usepackage{color}
\usepackage{color,soul}
\usepackage{verbatim}
\usepackage{graphicx}
\usepackage{cite}
\usepackage{amssymb}
\definecolor{LightBlue}{rgb}{0.75,0.936,1.00}
\sethlcolor{LightBlue}
\definecolor{Yellow}{rgb}{1.00, 1.00, 0.2}
\sethlcolor{Yellow}
\definecolor{LightCyan}{rgb}{0.88,1,1}

\begin{document}

\title{Performance of User-Assisted Nonlinear Energy Harvesting NOMA Network with Alamouti/MRC}

\author{Büşra Demirkol and Oğuz Kucur
\thanks{Büşra Demirkol and Oğuz Kucur are with the Department of Electronics Engineering, Gebze Technical University, Turkey (Email: \tt{\{busrademirkol, okucur\}}@gtu.edu.tr)}
\thanks{Manuscript received XXX XX, 20XX; revised XXXX XX, XXXX.}}

\markboth{To be submitted to IEEE Journals, Vol. XX, No. XX, XXXX 2024}%
{Shell \MakeLowercase{\textit{et al.}}: Bare Demo of IEEEtran.cls for IEEE Journals}

\IEEEpubid{0000--0000/00\$00.00~\copyright~2024 IEEE}

\IEEEpubidadjcol
\maketitle
\begin{abstract}
This paper focuses on evaluating the outage performance of a dual-hop single-phase non-orthogonal multiple-access (NOMA) system. The base station employs the Alamouti space-time block coding technique (Alamouti-STBC), enabling simultaneous communication with two mobile users, and the far user employs a maximal ratio combining (MRC) scheme. In this setup, the near user serves as a full-duplex (FD) (or half-duplex (HD)) energy harvesting (EH) relay, adopting decode-and-forward (DF) protocol for the far user. The study involves the development of a system model and the closed-form equations of exact and asymptotic outage probabilities (OP) over Nakagami-$m$ fading channels with and without direct link considering a threshold-based nonlinear EH relaying model. We verify analytical results by Monte Carlo simulations and show that the presence of a direct link in the system enhances the performance of the far user considerably by mitigating the degradation caused by the self-interference in the near user. 
\end{abstract}

\begin{IEEEkeywords}
Non-orthogonal multiple access, nonlinear energy harvesting, Alamouti, maximum ratio combining, outage probability.
\end{IEEEkeywords}

\section{Introduction}
\IEEEPARstart{N}{on}-orthogonal multiple access (NOMA) techniques, which have attracted great attention with their promising applications in the 5th generation and above (5G+) networks, provide simultaneous transmission by allowing the use of all resources accessible to multiple users, unlike traditional orthogonal multiple access (OMA) \cite{Aldababsa2018}. 
In power domain-based NOMA, where different power levels are allocated to multiple users inversely proportional to channel conditions, a superimposed signal is transmitted to users according to the superposition principle, and successive interference cancellation (SIC) is applied at the receivers. This ensures a good trade-off between system efficiency and user fairness \cite{Aldababsa2018}. In \cite{Ding2015}, a new NOMA scheme, called cooperative NOMA was proposed, where users with better channel conditions are utilized to transmit information to users with poor channel conditions. In this study, outage probability (OP) and diversity order (DO) were analyzed, and based on these analytical results, it was demonstrated that cooperative NOMA can provide maximum diversity gain for all users.

In addition to the primary motivation of improving spectral efficiency, enhancing energy efficiency is also crucial for 5G+ networks. Simultaneous wireless information and power transfer (SWIPT) was proposed to achieve both information and energy transfer through the same radio frequency signals simultaneously, aiming for more energy-efficient networks \cite{Ozyurt2022}. 
Most SWIPT studies focus on the linear energy harvesting (EH) model \cite{Nasir2013, Ding2014b, Liu2016, Ashraf2017, Xu2017, Demirkol2023, Alsaba2018, Wu2019, Huang2019, Aswathi2021, Guoan2023, Qiuyan2022}, while fewer investigate the practical nonlinear EH model \cite{Ma2021, Baranwal2023, Khennoufa2023, Bakshi2024}.
In \cite{Nasir2013}, a dual-hop network has been proposed, where an energy-constrained amplify-and-forward (AF) relay node harvests energy from the received signal using time switching (TS) and power splitting (PS) protocols and utilizes it to transmit information from the source to the destination. 
On the other hand, in \cite{Ding2014b}, the application of SWIPT to networks with randomly located decode-and-forward (DF) relays has been investigated using stochastic geometry in a cooperative scenario with multiple source nodes and a single destination. 
\cite{Liu2016} has explored the implementation of SWIPT in NOMA networks where users are randomly positioned. In this study, a novel cooperative SWIPT-NOMA protocol has been proposed, where EH relays act to assist weaker NOMA users by serving as energy harvesters.
The authors in \cite{Ashraf2017} have introduced the SWIPT-NOMA network where strong user, acting as PS-EH half-duplex (HD) relay, uses the harvested energy from the first time slot to transmit messages of weak users in the second time slot. The base station (BS) and strong user equipped with multi-antenna use beamforming and random selection, while the weak user has only a single antenna.
In \cite{Xu2017}, a  multi-input single-output (MISO) NOMA network, in which the strong user acts as a PS-EH HD relay to assist the weaker user, has been proposed. The system maximizes the data rate of the strong user while ensuring the service quality requirement for the weak user. 
The authors of \cite{Demirkol2023} investigates the OP of a dual-hop NOMA network, where the BS using Alamouti-space time block coding serves multiple users employing maximal ratio combining (MRC) via an EH-HD AF relay with single antenna. 
\IEEEpubidadjcol
In \cite{Alsaba2018}, a cooperative NOMA combined with beamforming network has been analyzed in terms of OP and sum-rate, where a user near to the source serves as a full-duplex (FD)-EH relay for a far user.
In \cite{Wu2019}, the maximization of the data rate for a near user serving as an FD-EH relay with multiple antennas has been studied in a similar system where the BS has a single antenna, unlike the one in \cite{Alsaba2018}.
\cite{Huang2019} has focused on the cooperative NOMA system where transmission from the BS to the near user is direct, and to the far user is achieved through an FD relay. The energy efficiency of this system is optimized based on PS ratios and beamforming vectors. 
In \cite{Aswathi2021}, the OP and efficiency performances of a cooperative NOMA system using TS-based SWIPT and FD relay have been investigated under the imperfect SIC effect, considering both direct and indirect connections from the BS to the far user.
\cite{Guoan2023} and \cite{Qiuyan2022} study reconfigurable intelligent surface aided cooperative transmission in two-user MISO SWIPT-NOMA networks.
Since EH circuits contain nonlinear elements such as diodes, transistors and capacitors, some nonlinear EH models that are more suitable for practice have been defined for harvested energy \cite{Boshkovska2015, Pejoski2018}.
Therefore, different from the aforementioned works using linear EH model, the nonlinear EH models are considered in \cite{Ma2021, Baranwal2023, Khennoufa2023, Bakshi2024}.  
\cite{Ma2021} investigates the OP of FD cooperative nonlinear EH NOMA system, where the near user relays transmission to the far user. It presents an optimal power allocation factor that can be adjusted without channel state information, and shows that self-recycled energy and self-interference (SI) improve OP. \cite{Baranwal2023} analyzes the OP of a cooperative FD-DF NOMA scheme with nonlinear EH, deriving closed-form expressions for user and system outage, in which a near user acts as a DF relay to serve far users and all nodes have single-antenna. The authors of \cite{Khennoufa2023} analyzes the OP, ergodic capacity, and throughput performances of an harvest-then-cooperate assisted NOMA system with linear and nonlinear EH, considering practical impairments. 
\cite{Bakshi2024} investigates cooperative SWIPT NOMA with battery-assisted nonlinear EH. Using PS-FD relaying, with and without a direct link, the study derives OP and throughput expressions considering the impacts of imperfect SIC and interference. 

Although the works mentioned above have extensively analyzed NOMA, SWIPT, multi-antenna systems, and FD relaying individually, more research is still needed on their comprehensive integration. 
Moreover, even Alamouti with EH-HD relaying has not been studied for user-assisted relaying case.
Different from the above papers, the objective of the current work is to analyze the OP of a two-user downlink EH-NOMA network using Alamouti/MRC under PS, whose simpler form examined by simulation in \cite{Demirkol2023b}. Specifically, the proposed system involves a source (BS) with multiple transmit antennas simultaneously sending signals to NOMA users, where the near user acts as an EH-FD relay for the far one equipped with multiple receive antennas. 
In this way, with FD relay the system can operate in single-phase. 
In this battery-assisted NOMA network, we utilize the piecewise nonlinear EH model in \cite{Pejoski2018}, which is more suitable for practice due to the limitations of IoT networks.
Leveraging the derived signal-to-interference+noise ratio (SINR) statistics, we formulate closed-form expressions for the OP of the considered system. 

This paper is organized as follows. The system model, performance analysis, numerical results and conclusion are given in Sections II, III, IV and V, respectively.

\section{System Model}

\begin{figure}[!t]
\centering
\includegraphics[width=3.5in]{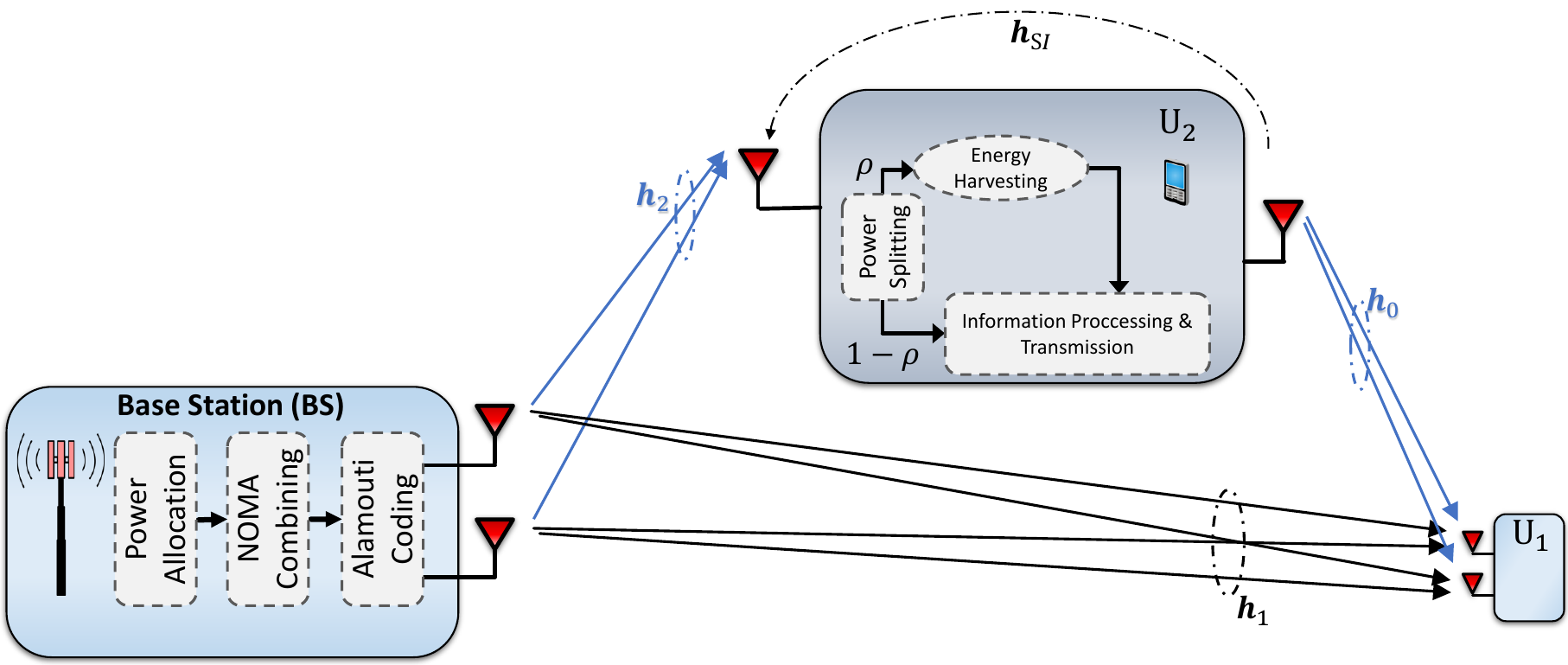}
\caption{System Model.}
\label{fig1}
\end{figure}
In this study, a single-phase point-to-point NOMA system is investigated, in which the near user operates in FD (or HD) mode to transmit from the BS to the far user, as depicted in Fig. \ref{fig1}. 
This system can be an application of local short-range communication, such as Bluetooth and ultra-wideband, as in \cite{Liu2016}.  
We mean by single-phase that the system works like a single-hop system due to user-assisted FD relay.
User 2 ($U_2$) utilizes a DF protocol to decode and forward information to User 1 ($U_1$). 
In the system, $U_2$ is equipped with one transmit and one receive antenna for implementing FD communication, $U_1$ is equipped with $N$ antennas to apply MRC. 
It is assumed that all wireless links in the network experience independent identically distributed (i.i.d.) Nakagami-$m$ fading. 
Due to its operation in FD mode, $U_2$ is subjected to imperfect SI. 
SI is modeled by a Gaussian random variable denoted as $h_{SI}$, with average power $\Omega_{SI} = E[|h_{SI}|^2] = \sigma_{SI}^2$, where we assume that $U_2$, acting as a relay, is exposed to residual SI effects arising from active and/or passive SI cancellation techniques \cite{Duarte2012}. At the $n$th time interval, $U_2$ simultaneously receives the NOMA signal from the BS and the SI signal. Due to Alamouti processing at the BS, the Alamouti code matrix $G_2$ representing the combined NOMA signals according to the superposition principle is expressed as $\textbf{G}_2 = \begin{bmatrix} s_1 & s_2 \\ -s_2^* & s_1^* \end{bmatrix}$. Here, the first and second rows of the matrix correspond to the signals to be transmitted in the first and second time intervals, respectively, and $\left(^*\right)$ denotes the complex conjugate. $s_1 = \sqrt{\frac{a_1 P_S}{2}} x_{11} + \sqrt{\frac{a_2 P_S}{2}} x_{21}$ and $s_2 = \sqrt{\frac{a_1 P_S}{2}} x_{12} + \sqrt{\frac{a_2 P_S}{2}} x_{22}$ respectively represent the combined NOMA signals transmitted from the first and second antennas in the first time interval according to the superposition principle. Here, $x_{lt}$ $\left(l \in \{1,2\}\right)$ and $\left(t \in \{1,2\}\right)$  represents the $t$th unit energy symbols of the $l$th user, where $a_l$ denotes the power allocation coefficient for the $l$th user, and $P_S$ represents the transmission power at the BS. Accordingly, the received signal at $U_2$ is expressed as
\begin{equation}
	\label{Equation1}
         \textbf{y}_2[n] = \textbf{G}_2[n] \textbf{h}_{\text{2}} + \varpi \textbf{h}_{\text{SI}} \sqrt{P_R} \tilde{x}_1 [n-D] + \textbf{n}_2[n],
\end{equation}
where $\textbf{h}_{\text{2}} = \left[h_{2i}\right]_{2\times1}$ is the channel coefficient vector between the BS and $U_2$, and $\Omega_2 = E[|h_{2i}|^2] = d_{S2}^{-\varepsilon}$, where $i$, $\varepsilon$ and $E[.]$ denotes the index of transmit antenna, path loss exponent and expectation operator, respectively. $\varpi$ denotes relay mode where $\varpi = 1$ and $\varpi = 0$ represent FD and HD mode, respectively. $\tilde{x}_1 [n-D]$ is the decoded signal of information transmitted from $U_2$ to $U_1$, where $D$ $\left(D \geq 1\right)$ represents the processing delay at $U_2$. $P_R$ denotes the average power harvested at $U_2$ and expended during the transmission to $U_1$. According to the PS-SWIPT protocol, the power of the received signal is divided into portions of $\rho:1-\rho$ ratio, where $\rho $ is PS ratio. $\rho P_S$ is utilized for EH, while $\left(1-\rho\right) P_S$ is used for information decoding (ID). Consequently, the signal vectors for EH and ID can be expressed as $\textbf{y}_{2}^{EH}[n] = \sqrt{\rho} \textbf{y}_{2}[n]$ and $\textbf{y}_{2}^{ID}[n] = \sqrt{1-\rho} \textbf{y}_{2}[n] = \sqrt{1-\rho} \left(\textbf{G}_2[n] \textbf{h}_{\text{2}} + \varpi \textbf{h}_{\text{SI}} \sqrt{P_R} \tilde{x}_1 [n-D]\right) + \tilde{\textbf{n}}_2[n] $, respectively. 
Here, $\tilde{\textbf{n}}_2[n] = (1-\rho)\textbf{n}_2[n]$ is the noise vector whose elements are complex Gaussian distributed with $CN(0, \sigma^2)$. 
Since, we adopt the nonlinear EH model in\cite{Pejoski2018}, the amount of the harvested power can be expressed with a piecewise linear function as $ P_R = \begin{cases} max\{0,\frac{\eta\rho P_{S}\|\textbf{h}_{\text{2}}\|_F^2}{1-\eta\rho\varpi\sigma_{SI}^2}\}, & P_{in}\leq P_{th}\\ \eta P_{th}, & P_{in} > P_{th}\end{cases} $. Here, the $max\{\cdot,\cdot\}$ operator ensures that the harvested power remains within a meaningful range, $\eta$ is energy conversion efficiency and $P_{in}=\rho\left(P_S\|\textbf{h}_{\text{2}}\|_F^2+\varpi P_R \sigma_{SI}^2\right)$ is the total power available for EH in $U_2$. According to the NOMA protocol, $U_2$ employs SIC to decode its own signal and the signal from $U_1$. The expression for the SINR indicating the instantaneous signal that $U_2$ decodes with respect to $U_1$ can be given as $\gamma_{2,1} = \frac{a_1\frac{ P_S}{2} (1-\rho)  \|\textbf{h}_{\text{2}}\|_F^2}{a_2\frac{ P_S}{2}(1-\rho) \|\textbf{h}_{\text{2}}\|_F^2 + \varpi(1-\rho)\sigma_{SI}^2 P_R+ \sigma^2}$, where $\|.\|_F$ denotes Frobenius norm. The instantaneous SINR expression indicating that $U_2$ decodes its own signal can be obtained as $\gamma_{2,2} = \frac{a_2\frac{ P_S}{2}(1-\rho) \|\textbf{h}_{\text{2}}\|_F^2}{\varpi(1-\rho)\sigma_{SI}^2 P_R+ \sigma^2}$.
In FD mode, after decoding the information signal, $U_2$ forwards $\tilde{x}_1[n]$ to $U_1$. Subsequently, $U_1$ employs an MRC receiver to combine and decode the signals received from both the BS and $U_2$ \cite{Fidan2020}. Accordingly, the received signal at $U_1$ is given by:
\begin{equation}
	\label{Equation4}
         \textbf{y}_1[n] = \textbf{G}_2[n] \textbf{h}_{\text{1}} + \sqrt{P_R} \textbf{h}_{\text{0}}\tilde{x}_1[n-D] + \textbf{n}_1[n] 
\end{equation}
Here, $\textbf{h}_{\text{1}}$ represents the channel coefficient vector between the BS and $U_1$, while $\textbf{h}_{\text{0}}$ denotes the channel coefficient vector between $U_2$ and $U_1$. The channel power gains are given by $\Omega_1 = E[|h_{1i}|^2] = d_{S1}^{-\varepsilon}$ and $\Omega_0 = E[|h_{0t}|^2] = d_{21}^{-\varepsilon}$. The noise vector $\textbf{n}_1[n] = [n_{1}^t]_{2\times N}$ consists of elements with complex Gaussian distribution $\left(CN(0, \sigma^2)\right)$. It is assumed that the signals from the BS and $U_2$ to $U_1$ can be completely resolved, allowing for proper phase alignment and combination by MRC \cite{Fidan2020}. Accordingly, the instantaneous SINRs for the signals from the BS and $U_2$ to $U_1$ are obtained as $\gamma_{1,1} = \frac{a_1 \frac{P_S}{2} \|\textbf{h}_{\text{1}} \|_F^2}{a_2 \frac{P_S}{2} \|\textbf{h}_{\text{1}} \|_F^2 + \sigma^2}$ and $\gamma_{1,2} = \frac{P_R \|\textbf{h}_{\text{0}}\|_F^2}{\sigma^2}$, respectively. 
In this scenario, the signals directly from the BS and from $U_2$ are combined at $U_1$ using MRC. As a result, the SINR received at $U_1$ after MRC, denoted as $\gamma_{1}$, is given by the sum of the SINRs obtained previously: $\gamma_{1} = \gamma_{1,1} + \gamma_{1,2}$.

Since the magnitudes of the fading gains are assumed to follow Nakagami-$m$ distributions, $\|\textbf{h}_{X}\|_F^2$ $\left(X\in \{1,2,0\}\right)$ follows a Gamma distribution, and its cumulative distribution function (CDF) and probability density function (PDF) are $F_{\|\textbf{h}_{X}\|_F^2}(x) = 1 - e^{-\frac{xm}{\Omega_X}} \sum_{n=0}^{mN-1} \frac{1}{n!} \left(\frac{xm}{\Omega_X}\right)^n$ and $f_{\|\textbf{h}_{\text{X}}\|_F^2}(x) =  \frac{\left(\frac{m}{\Omega_1}\right)^{2mN} x^{2mN-1}}{\Gamma(2mN)} e^{-\frac{xm}{\Omega_X}}$, respectively, where $N=1$ for $X=2$. 

\section{Performance Analyses}
In this section, OP is derived. It can be defined as the probability of the instantaneous SINR for a given user falling below a specific threshold value, i.e., $\gamma_{th_l} = 2^{R_l} - 1$ for FD mode and  $\gamma_{th_l} = 2^{2R_l} - 1$ for HD mode, where $R_l$ represents the bits per channel in use (BPCU) for the $l$th user.
The OP will be evaluated in two representative scenarios hereafter.
\subsection{User-Assisted Without Direct Link}
\subsubsection{OP Analysis}
According to the NOMA, the outage events for $U_2$ can be explained as follows: $U_2$ cannot decode the message related to $U_1$ and, consequently, its own message. Therefore, the OP for U2 can be expressed as $P_2 = 1 - P(\gamma_{2,1} > \gamma_{\text{th}_1}, \gamma_{2,2} > \gamma_{\text{th}_2}) = 1 - P(P_{in}\leq P_{th}, \gamma_{2,1} > \gamma_{\text{th}_1}, \gamma_{2,2} > \gamma_{\text{th}_2})-P(P_{in} > P_{th}, \gamma_{2,1} > \gamma_{\text{th}_1}, \gamma_{2,2} > \gamma_{\text{th}_2}) = 1-I_1-I_2$. By substituting $\gamma_{2,1}$ and $\gamma_{2,2}$ into $P_2$, we can express $I_1=P\left(\tau_1^*<\|\textbf{h}_{\text{2}}\|_F^2<\beta_1\right) = F_{\|\textbf{h}_{\text{2}}\|_F^2}(\beta_1)-F_{\|\textbf{h}_{\text{2}}\|_F^2}(\tau_1^*)$ under the condition of $\beta_1>\tau_1^*$, otherwise $I_1 = 0$, and $I_2 = P\left(\|\textbf{h}_{\text{2}}\|_F^2>\beta_1^*\right) =1- F_{\|\textbf{h}_{\text{2}}\|_F^2}(\beta_1^*)$, where $\tau_1^* = \max( 0,\tau_{1}, \tau_{2})$, $\tau_{1} \triangleq  \frac{2\gamma_{\text{th}_1}}{\frac{P_S}{\sigma^2}(1-\rho)(a_1 - (a_2+2\phi_1) \gamma_{\text{th}_1})}$, and $\tau_{2} \triangleq \frac{2\gamma_{\text{th}_2}}{\frac{P_S}{\sigma^2}(1-\rho)(a_2 - 2\phi_1 \gamma_{\text{th}_2})}$. $\beta_1^*= \max(\beta_{1},\beta_{2},\beta_{3})$ and $\beta_1  \triangleq \frac{P_{th} \left(1-\eta\rho\varpi\sigma_{SI}^2 \right)}{\rho P_S}$, $\beta_2  \triangleq \frac{2 \gamma_{th_1} \left(\sigma^2 + \eta (1-\rho) \varpi \sigma_{SI}^2 P_{th} \right)}{ P_S (1-\rho)(a_1 - a_2 \gamma_{\text{th}_1})}$, and $\beta_3  \triangleq \frac{2 \gamma_{th_2} \left(\sigma^2 + \eta (1-\rho) \varpi \sigma_{SI}^2 P_{th} \right)}{a_2 P_S (1-\rho)}$. $P_2$ is obtained under the condition $a_1 > a_2 \gamma_{\text{th}_1}$ and $a_1 > (a_2+ 2\phi_1) \gamma_{\text{th}_1}$, where $\phi_1 = \frac{\eta\rho\varpi \sigma_{SI}^2}{1-\eta\rho \varpi \sigma_{SI}^2}$. Thus, the OP of $U_2$ can be obtained using the CDF of $\|\textbf{h}_{\text{2}}\|_F^2$ at the value of $\tau_1^*$, $\beta_1^*$ and $\beta_1$.
On the other hand, there are two possible outage events for $U_1$. Firstly, when $U_2$ fails to detect its own signal. Secondly, when $U_1$ cannot detect its own signal under the condition of $U_2$ successfully detecting its own signal. Then, the OP of $U_1$ is expressed as: 
\begin{equation}
    \label{Equation10}
    \begin{split}
    P_{1} &= P (\gamma_{2,1} < \gamma_{\text{th}_1}) + P (\gamma_{1,2} < \gamma_{\text{th}_1}, \gamma_{2,1} > \gamma_{\text{th}_1})\\ 
          &= \Upsilon_1 + \Upsilon_2 + \Upsilon_3 + \Upsilon_4,
    \end{split}
\end{equation}
where $\Upsilon_1 = P (P_{in} \leq P_{th}, \gamma_{2,1} < \gamma_{\text{th}_1})$, $\Upsilon_2 = P(P_{in} > P_{th}, \gamma_{2,1} < \gamma_{\text{th}_1})$, $\Upsilon_3 = P(P_{in}\leq P_{th},\gamma_{2,1} < \gamma_{\text{th}_1}, \gamma_{1,1} < \gamma_{\text{th}_1})$ and $\Upsilon_4 = P(P_{in}> P_{th},\gamma_{2,1} < \gamma_{\text{th}_1}, \gamma_{1,1} < \gamma_{\text{th}_1})$. By substituting $\gamma_{2,1}$, and $\gamma_{1,2}$ into $\left(\ref{Equation10}\right)$, the expressions of $\Upsilon_1$, $\Upsilon_2$, $\Upsilon_3$ and $\Upsilon_4$ in closed form can be calculated. Accordingly, $\Upsilon_1 = P(0<\|\textbf{h}_{\text{2}}\|_F^2<\tau_2^*) = F_{\|\textbf{h}_{\text{2}}\|_F^2}(\tau_2^*) $, where $\tau_2^* = min\{\tau_1, \beta_1\}$, and $\Upsilon_2 = P(\beta_1<\|\textbf{h}_{\text{2}}\|_F^2<\beta_2) = F_{\|\textbf{h}_{\text{2}}\|_F^2}(\beta_2)- F_{\|\textbf{h}_{\text{2}}\|_F^2}(\beta_1)$. Then, $\Upsilon_3$ can be expressed as 
\begin{equation}
    \label{Equation11}
    \begin{split}
    \Upsilon_3 &= P(\tau_3^*<\|\textbf{h}_{\text{2}}\|_F^2<\beta_1, \|\textbf{h}_{\text{0}}\|_F^2<\frac{\gamma_{\text{th}_1}}{\phi_2 P_S \|\textbf{h}_{\text{2}}\|_F^2})\\
               &= \int_{\tau_3^*}^{\beta_1} f_{\|\textbf{h}_{\text{2}}\|_F^2}(x) F_{\|\textbf{h}_{\text{0}}\|_F^2}\left(\frac{\gamma_{\text{th}_1}}{\phi_2 P_Sx} \right) dx
    \end{split}
\end{equation}
where $\tau_3^* = max\{0,\tau_1\}$ and $ \phi_2 = \frac{\eta\rho}{\sigma^2 (1-\eta\rho\varpi\sigma_{SI}^2)}$. By substituting CDF of $\|\textbf{h}_{\text{0}}\|_F^2$ of and PDF of $\|\textbf{h}_{\text{2}}\|_F^2$ into $\left(\ref{Equation11}\right)$, $\Upsilon_3$ can be obtained as follows: 
\begin{equation}
    \label{Equation12}
    \begin{split}
    \Upsilon_3 &=\left(F_{\|\textbf{h}_{\text{2}}\|_F^2}(\beta_1)-F_{\|\textbf{h}_{\text{2}}\|_F^2}(\tau_3^*)\right) \frac{\left(\frac{m}{\Omega_2}\right)^{2m}}{\Gamma(2m)}\sum_{n=0}^{mN-1}\sum_{p=0}^{\infty}       \frac{(-1)^p}{n!p!} \\
               &\times \left(\frac{m \gamma_{\text{th}_1}}{\phi_2 P_S\Omega_0}\right)^{n} \left(\frac{m}{\Omega_2}\right)^{p}\left(\frac{m \gamma_{\text{th}_1}}{\phi_2 P_S\Omega_0}\right)^{-\varphi_1 }\\
               &\times \left( \Gamma(\varphi_1 , \frac{m \gamma_{\text{th}_1}}{\phi_2 P_S\Omega_0 \beta_1})- \Gamma(\varphi_1, \frac{m \gamma_{\text{th}_1}}{\phi_2 P_S\Omega_0 \tau_3^*}) \right)
    \end{split}
\end{equation}
where $\varphi_1 = n-p-2m$ and $\Gamma(\cdot,\cdot)$ is the upper incomplete Gamma function. Finally, $\Upsilon_4$ is obtained as $\Upsilon_4 =P(\|\textbf{h}_{\text{2}}\|_F^2 >\beta_3^*, \|\textbf{h}_{\text{0}}\|_F^2<\frac{\gamma_{\text{th}_1}\sigma^2}{\eta P_{th}}) = (1- F_{\|\textbf{h}_{\text{2}}\|_F^2}(\beta_3^*) ) F_{\|\textbf{h}_{\text{0}}\|_F^2}(\frac{\gamma_{\text{th}_1}\sigma^2}{\eta P_{th}}) $.
\subsubsection{Asymptotic Analysis}
In this section, we derive asymptotic expressions for the OPs of the considered system to observe high SNR behaviors. In the high SNR region ($ P_S \rightarrow \infty $), the approximated OP expression for $U_2$ can be obtained as $P_{2}^{\infty} \approx F_{\|\textbf{h}_{\text{2}}\|_F^2}^{\infty}(\beta_1^*)+F_{\|\textbf{h}_{\text{2}}\|_F^2}(\tau_1^*)-F_{\lVert \mathbf{h}_2 \rVert_F^2}^{\infty}(\beta_1) = \frac{\left(m/\Omega_2\right)^{2m}}{\Gamma\left(2m+1\right))} \left( \left( \tilde{\beta}_1^*\right)^{2m} + \left( \tilde{\tau}_1^*\right)^{2m} - \left( \tilde{\beta}_1\right)^{2m}\right) \left(P_S\right)^{-2m}$, where $\tilde{\beta}_1^* = P_S \beta_1^*$, $\tilde{\tau}_1^* = P_S \tau_1^*$, and $\tilde{\beta}_1 = P_S \beta_1$. Thus, the DO is $2m$. When the transmit power $P_S$ approaches infinity ($ \bar\gamma \rightarrow \infty $), the CDF of $\lVert h_X \rVert_F^2$ can be asymptotically expressed as $F_{\lVert h_X \rVert_F^2}^{\infty}(x) = \frac{\left(\frac{xm}{\Omega_X}\right)^{mN}}{\Gamma(mN + 1)}$, where, the asymptotic property of the Gamma function $\gamma(v, x \rightarrow 0) \approx \frac{x^{v}}{\Gamma(v)} $ is employed \cite[eq.(45:9:1)]{Oldham2008}. Then, we can evaluate the asymptotic OP related to $U_2$. 
On the other hand, in the high SNR region, the expression of OP in $\left(\ref{Equation10}\right)$ can be approximated as  $P_1^\infty \approx F_{\|\mathbf{h}_2\|_F^2}^\infty(\tau_{2}^*) + \left(F_{\|\mathbf{h}_2\|_F^2}^\infty(\beta_{2}) - F_{\|\mathbf{h}_2\|_F^2}^\infty(\beta_{1})\right) + F_{\|\mathbf{h}_0\|_F^2}^\infty(\frac{\gamma_{\text{th}_1}\sigma^2}{\eta P_{th}})[1 - F_{\|\mathbf{h}_2\|_F^2}^\infty(\beta_{3}^*)] + \Upsilon_3^\infty $. Here, $\Upsilon_3^\infty =\int_{\tau_3^*}^{\beta_1} f_{\|\textbf{h}_{\text{2}}\|_F^2}^\infty(x) F_{\|\textbf{h}_{\text{0}}\|_F^2}^\infty\left(\frac{\gamma_{\text{th}_1}}{\phi_2 P_Sx} \right) dx$. By substituting approximated CDF and PDF into $\Upsilon_3^\infty$, we can obtain that $\Upsilon_3^\infty = \frac{2m \left(\frac{m}{\Omega_2}\right)^{2m} \left(\frac{m\gamma_{\text{th}_1} }{\phi_2 P_S \Omega_0}\right)^{mN} }{\Gamma(2m+1) \Gamma(mN+1)} \begin{cases}
ln(\beta_1) - ln(\tau_3^*), &  \nu = 0,\\
\frac{ \beta_1^{ \nu} - (\tau_3^*)^{ \nu} }{ \nu}, &   \nu \neq 0
\end{cases}$, where $\nu = m(2-N)$. Finally, we can obtain the asymptotic OP $P_{1}^{\infty}$ by substituting the related approximated CDFs and $\Upsilon_3^\infty$. Accordingly, $P_{1}^{\infty} = \begin{cases} (A+B+D) P_S^{-2m} + C P_S^0 + E P_S^{-mN},&  \nu=0,\\
(A+B+D+F) P_S^{-2m} + C P_S^0, & \nu \neq 0  \end{cases}$, where $A = \frac{\left(\frac{m}{\Omega_2}\right)^{2m} (\tau_2^* P_S)^{2m}}{\Gamma(2m+1)}$, $B = \frac{\left(\frac{m}{\Omega_2}\right)^{2m} ((\tau_1 P_S)^{2m} - (\beta_2 P_S)^{2m})}{\Gamma(2m+1)}$, $C = \frac{\left(\frac{m}{\Omega_0}\right)^{mN} (\frac{\gamma_{\text{th}_1}\sigma^2}{\eta P_{th}})^{mN}}{\Gamma(mN+1)} $, $D = \frac{C\left(\frac{m}{\Omega_2}\right)^{2m} (\beta_3^* P_S)^{2m}}{\Gamma(2m+1)} $, $E = \frac{2m \left(\frac{m}{\Omega_2}\right)^{2m} \left(\frac{m\gamma_{\text{th}_1} }{\phi_2 \Omega_0}\right)^{mN} }{\Gamma(2m+1) \Gamma(mN+1)}ln(\frac{\beta_1}{\tau_3^*})$ and $F =\frac{2m \left(\frac{m}{\Omega_2}\right)^{2m} \left(\frac{m\gamma_{\text{th}_1} }{\phi_2 \Omega_0}\right)^{mN} }{\Gamma(2m+1) \Gamma(mN+1)} \frac{ (\beta_2 P_S)^{\nu} - (\tau_3^* P_S)^{\nu}}{\nu}$. Thus, in the case of $\nu=0$ (i.e., $N=2$), the DO is $min\{0,2m,mN\} = 0$, and otherwise it is also $min\{2m,0\} = 0$.
\subsection{User-Assisted With Direct Link}
In this subsection, we examine a more demanding scenario wherein the direct link between the BS and $U_1$ is utilized for information transmission, potentially enhancing system reliability. Nevertheless, the OP of $U_2$ remains unaffected by this direct link. Therefore, we only present the exact and asymptotic OP of $U_1$ herein.
\subsubsection{OP Analysis}
Outage events for $U_1$ can be explained as follows: $U_2$ can successfully decode the message from $U_1$; however, after MRC at $U_1$, the SINR may fall below the threshold SINR. Alternatively, neither $U_2$ nor $U_1$ can decode the message from $U_1$. Therefore, the OP for $U_1$ can be expressed as follows:
\begin{equation}
    \label{Equation15}
\begin{split}
P_1 &= P \left(\gamma_{1} < \gamma_{\text{th}_1}, \gamma_{2,1} > \gamma_{\text{th}_1}\right) + P \left(\gamma_{2,1} < \gamma_{\text{th}_1}, \gamma_{1,1} < \gamma_{\text{th}_1}\right) \\  &= \chi_1++\chi_2+\chi_3+\chi_4
\end{split}
\end{equation}
Here, $\chi_1 = P(P_{in}\leq P_{th}, \gamma_{1} < \gamma_{\text{th}_1}, \gamma_{2,1} > \gamma_{\text{th}_1})$, $\chi_2 = P(P_{in} > P_{th}, \gamma_{1} < \gamma_{\text{th}_1}, \gamma_{2,1} > \gamma_{\text{th}_1})$, $\chi_3 = P(P_{in}\leq P_{th},\gamma_{2,1} < \gamma_{\text{th}_1}, \gamma_{1,1} < \gamma_{\text{th}_1})$ and $\chi_4 = P(P_{in} > P_{th},\gamma_{2,1} < \gamma_{\text{th}_1} ,\gamma_{1,1} < \gamma_{\text{th}_1})$. By substituting $\gamma_{2,1}$, $\gamma_{1}$ and $\gamma_{1,1}$ into $\left(\ref{Equation15}\right)$, the expressions of $\chi_1$, $\chi_2$, $\chi_3$ and $\chi_4$ in closed form can be calculated. Accordingly, 
\begin{equation}
    \label{Equation16}
\begin{split}
\chi_1 =&  P\left(\beta_2^*<\|\textbf{h}_{\text{2}}\|_F^2<\beta_1, \|\textbf{h}_{\text{0}}\|_F^2 < \lambda_1, \|\textbf{h}_{\text{1}}\|_F^2 < \theta_1\right) \\
=& \int_{\beta_2^*}^{\beta_1} \int_0^{\theta_1} f_{\|\textbf{h}_{\text{2}}\|_F^2}(x) F_{\|\textbf{h}_{\text{0}}\|_F^2}\left(\lambda_1\right) f_{\|\textbf{h}_{\text{1}}\|_F^2}(y) \, dydx,
\end{split}
\end{equation}
where $\beta \triangleq \frac{a_1 \overline{\gamma}/2 y}{a_2 \overline{\gamma}/2 y + 1}$, $\theta_1 \triangleq \frac{2\gamma_{\text{th}_1}}{\overline{\gamma}(a_1 - a_2 \gamma_{\text{th}_1})}$ and $\lambda_1 \triangleq \frac{\left(\gamma_{\text{th}_1}-\beta \right)}{\phi_2 P_S x}$ is defined, and $\beta_2^* = max\{0, \beta_2\}$. Notice that, $\beta_1$ is always smaller than $\beta_2^*$, thereby $\chi_1 =0$.
Subsequently, $\chi_2$ is expressed as
\begin{equation}
    \label{Equation17}
\begin{split}
\chi_2 =&  P(\|\textbf{h}_{\text{2}}\|_F^2>\beta_2^*, \|\textbf{h}_{\text{0}}\|_F^2 < \frac{\sigma^2}{\eta P_{th}} \left(\gamma_{\text{th}_1}-\beta \right), \|\textbf{h}_{\text{1}}\|_F^2 < \theta_1) \\
=& \bar{F}_{\|\textbf{h}_{\text{2}}\|_F^2}(\beta_2^*) \underbrace{\int_0^{\theta_1}  F_{\|\textbf{h}_{\text{0}}\|_F^2}\left(\frac{\sigma^2}{\eta P_{th}} \left(\gamma_{\text{th}_1}-\beta \right)\right) f_{\|\textbf{h}_{\text{1}}\|_F^2}(y) \, dy}_{X_{21}},
\end{split}
\end{equation}
where $\bar{F}_{X}(\cdot) = 1-F_{X}(\cdot)$. Substituting the provided PDF and CDF expressions of $\|\textbf{h}_{\text{1}}\|_F^2$ and $\|\textbf{h}_{\text{0}}\|_F^2$ into $X_{21}$, and performing the variable transformation $a_2 \bar{\gamma}/2 y + 1 = u$, the equation can be expressed as follows:
\begin{equation}
   \label{Equation18}
    \begin{split}
X_{21} &= F_{\|\mathbf{h}_0\|_F^2}(\theta_1)-\frac{\left(\frac{2m}{a_2 P_S \Omega_1}\right)^{2mN}}{\Gamma(2mN)} e^{\frac{2m}{a_2 P_S \Omega_1}} e^{-\frac{m\sigma^2 ( \gamma_{\text{th}_1} -\frac{a_1}{a_2})}{\eta P_{th} \Omega_0}}\\ 
&\times\sum_{n,t,p,k} \binom{n}{k} \binom{2mN-1}{t} \frac{ (-\sigma^2)^{2mN-1-t}(-1)^{p} }{n! p!}\\ 
&\times\left(\frac{m\sigma^2}{\eta P_{th}\Omega_0}\right)^n   \left(\frac{2m}{a_2P_S \Omega_1}\right)^p \left(\gamma_{\text{th}_1} - \frac{a_1}{a_2}\right)^{n-k}\left(\frac{a_1\sigma^2}{a_2}\right)^k \\ 
&\times \underbrace{\int_{\sigma^2}^{a_2 P_S/2 \theta_1+\sigma^2} u^{t+p-k} e^{-\frac{m(\sigma^2)^2 a_1}{\eta P_{th} \Omega_0 a_2}\frac{1}{u}} \, du}_{I_3}
 \end{split}
\end{equation}
where $\sum_{n,t,p,k} = \sum_{n=0}^{mN-1}\sum_{k=0}^{n} \sum_{t=0}^{2mN-1} \sum_{p=0}^{\infty}$. $\left(\ref{Equation18}\right)$ is obtained using the Taylor series expansion. Subsequently, the variable transformation $u = \frac{1}{x}$ is applied, and employing \cite[eq.(3.381.3)]{Grad2000}, $I_3$ is expressed as follows:
\begin{equation}
    \label{Equation19}
    \begin{split}
    I_3 &= \lambda_2^{-\varphi_2}\left( \Gamma(\varphi_2,\lambda_2 \varepsilon_2)-\Gamma(\varphi_2,\lambda_2 \varepsilon_1 )\right)
 \end{split}
\end{equation}
where $\lambda_2 =\frac{m(\sigma^2)^2 a_1}{\eta P_{th}\Omega_0 a_2} $, $\varphi_2 = k-t-p-1$, $\varepsilon_1 = 1/\sigma^2$ and $\varepsilon_2 = 1/(a_2 \frac{P_S}{2} \theta_1 + \sigma^2)$ . Then, $\chi_3$ can be expressed as $\chi_3 = P(0<\|\textbf{h}_{\text{2}}\|_F^2<\tau_2^*, \|\textbf{h}_{\text{1}}\|_F^2 < \theta_1) = F_{\|\textbf{h}_{\text{2}}\|_F^2} (\tau_2^*)-F_{\|\textbf{h}_{\text{1}}\|_F^2} (\theta_1)$. Finally, $\chi_4$ is obtained as $\chi_4 = P(\beta_2<\|\textbf{h}_{\text{2}}\|_F^2<\beta_3, \|\textbf{h}_{\text{1}}\|_F^2 < \theta_1) = \left(F_{\|\textbf{h}_{\text{2}}\|_F^2} (\beta_3)-F_{\|\textbf{h}_{\text{2}}\|_F^2} (\beta_2)\right) F_{\|\textbf{h}_{\text{1}}\|_F^2} (\theta_1)$. The closed-form expression for the OP related to $U_1$ is obtained by substituting the expressions in $\chi_1=0$, $\left(\ref{Equation17}\right)$, $\left(\ref{Equation18}\right)$, $\left(\ref{Equation19}\right)$, $\chi_3$ and $\chi_4$ into $\left(\ref{Equation15}\right)$. 
\subsubsection{Asymptotic Analysis}
In this section, asymptotic analysis is provided for the case with direct link. In the high SNR regime ($P_S \rightarrow \infty$), an approximate expression for the OP of $U_1$ can be obtained as $P_{1}^{\infty} \approx  \chi_2^\infty + (F_{\lVert \mathbf{h}_2 \rVert_F^2}^{\infty}(\tau_2^*)+ F_{\lVert \mathbf{h}_2 \rVert_F^2}^{\infty}(\beta_3) -F_{\lVert \mathbf{h}_2 \rVert_F^2}^{\infty}(\beta_2))  F_{\lVert \mathbf{h}_1 \rVert_F^2}^{\infty}(\theta_1)$. Here, $\chi_2^\infty = (1 -F_{\lVert \mathbf{h}_2 \rVert_F^2}^{\infty}(\beta_2^*)) \int_{0}^{\theta_1} f_{\|\textbf{h}_{\text{1}}\|_F^2}^\infty(y) F_{\|\textbf{h}_{\text{0}}\|_F^2}^\infty\left(\lambda_1 \left(\gamma_{\text{th}_1}-\beta \right)\right) dx$. By substituting CDF and PDF into $\chi_2^\infty$ and after some mathematical manipulations, $\chi_2^\infty$
 can be obtained as follows:
\begin{equation}
    \label{Equation28}
    \begin{split}
\chi_2^\infty &= \frac{2mN\left(\frac{2m}{a_2 P_S \Omega_1}\right)^{2mN} \left(\frac{m\sigma^2}{\eta P_{th} \Omega_0}\right)^{mN}}{\Gamma(2mN+1) \Gamma(mN+1)}  \sum_{t=0}^{2mN-1}\sum_{p=0}^{mN}\\ 
              &\times \binom{2mN-1}{t} \binom{mN}{p} (-\sigma^2)^{2mN-1-t} \left(\frac{a_1}{a_2}\right)^{mN} \\
              &\times \left(\frac{\gamma_{\text{th}_1}\frac{a_1}{a_2}-1}{\sigma^2}\right)^{mN-p}   
              \begin{cases} ln(\varepsilon_1) - ln(\varepsilon_2), & \kappa = 1,
                \\ \frac{ \varepsilon_1^{ \kappa } - \varepsilon_2^{ \kappa } }{ \kappa }, &   \kappa  \neq 1 ,
                \end{cases}  
    \end{split}
\end{equation}
where $\kappa = p-t-1$. Accordingly, we can obtain $P_{1}^{\infty}$ by substituting the related approximated CDFs nd $\chi_2^\infty$. Then, similar to Section III.A.2, the DO can be found to be $2m(N+1)$.
\section{Numerical Results}
In this section, the OP performance of the system has been examined through theoretical analysis and simulation results. In the figures, the path loss exponent is set as $\varepsilon=2$, energy conversion efficiency as $\eta=0.7$ and the Nakagami-$m$ parameter as $m=1$. Power allocation coefficients assigned to the users are $a_1=0.8$ and $a_2=0.2$. The target rate values with and without direct link cases set to $R_{\text{1}}= 1, R_{\text{2}}=2$ and $R_{\text{1}}= 0.5, R_{\text{2}}=3$, respectively. It is assumed that the distance between the BS and $U_1$ is normalized, i.e., $d_{S1}=1.5$, $d_{S2}=1$, and $d_{21}=d_{S1}-d_{S2}=0.5$.

\begin{figure} 
	\centering
	\includegraphics[width=2.5in]{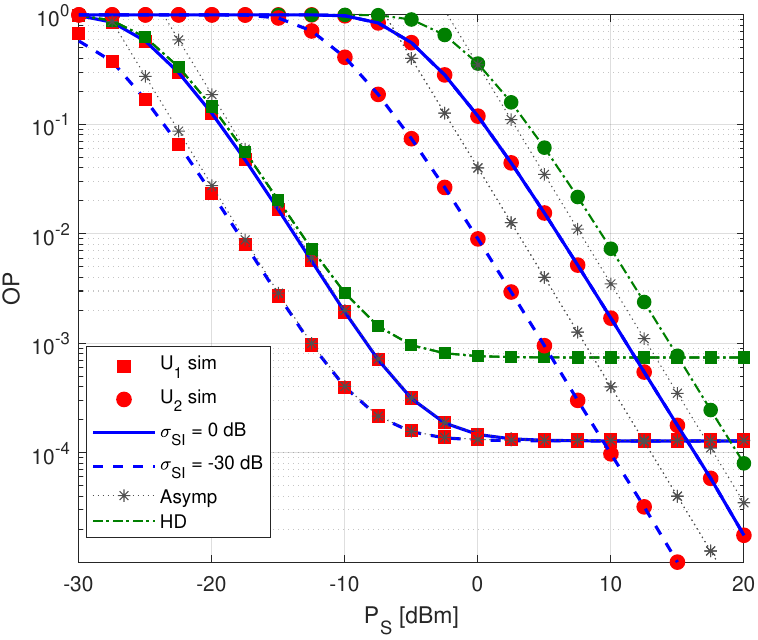}
	\caption{The OP performance of the system without direct link.}
	\label{fig5}
\end{figure}

The OP performance versus SNR of the system without direct link is presented in Fig. \ref{fig5} for PS ratio $\rho= 0.5$, number of receive antennas $N=2$ and different $\sigma_{SI}$ values. 
As seen, theoretical and simulation results are in perfect agreement. 
For all $\sigma_{SI}$ values, $U_1$ outperforms $U_2$ significantly up to $SNR=10$ dB. The OP of $U_1$ reaches error floor at $SNR = 0$ dB due to nonlinear EH.
Additionally, to increase performance for both users, $\sigma_{SI}$ should be kept at low values. 
At an OP of $10^{-3}$, the SNR gains for $U_1$ and $U_2$ are $4$ and $6$ dB, respectively as $\sigma_{SI}$ decreases from $0$ dB to $-30$ dB.
Thus, $\sigma_{SI}$ has a more significant effect on the performance of $U_2$ than $U_1$.
Moreover, comparing the FD and HD cases, for both $U_1$ and $U_2$ the performance of the FD case is better than that of the HD case for all SNR and $\sigma_{SI}$ values.
For $U_1$, the HD case is close to the FD case for $\sigma_{SI} = 0$ dB. As seen, DO of $U_1$ is $0$ for high SNR region validating the analysis while it is the same as that of $U_2$ for low SNR region, which is $2$.

\begin{figure} 
	\centering
	\includegraphics[width=2.5in]{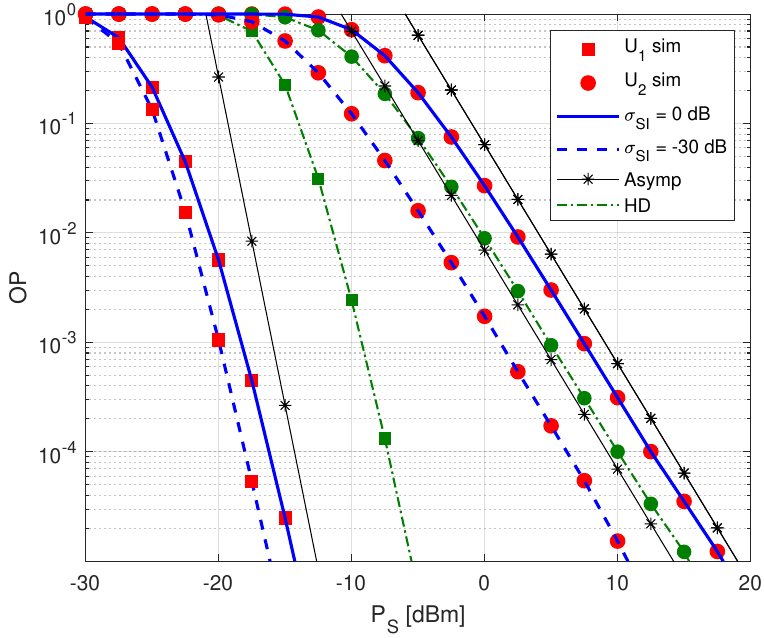}
	\caption{The OP performance of the system with direct link.}
	\label{fig2}
\end{figure}

Fig. \ref{fig2} presents the OP performance for the system with direct link for PS factor $\rho=0.5$, number of receive antennas $N=2$ and different $\sigma_{SI}$ values.
The results similar to those in Fig. \ref{fig5} are obtained, when the performances of users are compared for different values of $\sigma_{SI}$.
For $U_2$, the performance of the FD case is better than the HD case for $\sigma_{SI} = -30$ dB, while the performance of the HD case is better for $\sigma_{SI} = 0$ dB, different from the case without direct connection.
It is observed that having a direct link in the system improves the performance of $U_1$ by reducing the degradation caused by the SI in $U_2$ and causes the error floor subjected to nonlinear EH to disappear.

\begin{figure}
	\centering
	\includegraphics[width=2.5in]{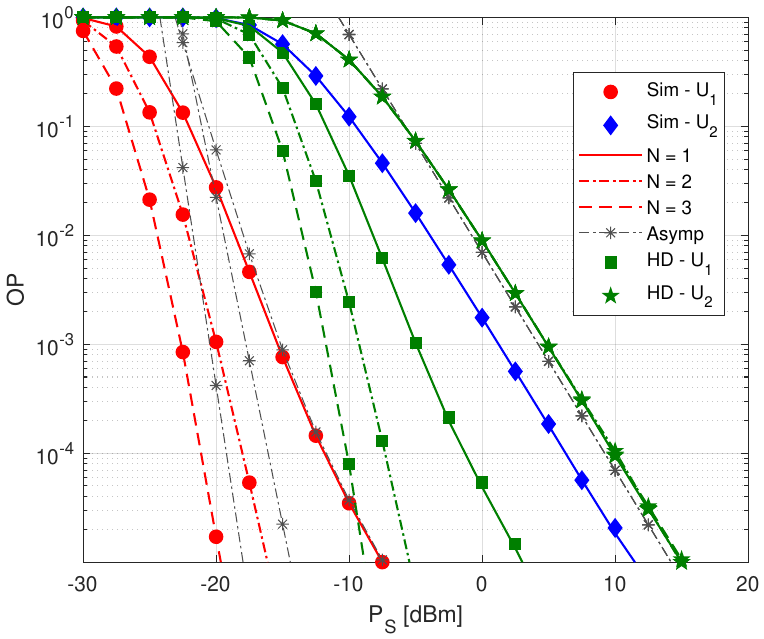}
	\caption{The OP of the system with different number of antennas.}
	\label{fig3}
\end{figure}

In Fig. \ref{fig3}, the OP performance of the system with direct link has been investigated for $\sigma_{SI}=-30$ dB and $\rho=0.5$ with different number of receive antennas. 
In the FD case, for an OP of $10^{-4}$ when the number of receive antennas of $U_1$ increases from $N=1$ to $N=2$ and from $N=2$ to $N=3$, approximately $4.6$ dB and $2.6$ dB  SNR gains are achieved, respectively. 
These SNR gains are $4.3$ dB and $2.45$ dB in the HD case.
Also, the figure confirms that the DO of the system is $2m$ and $2m(N+1)$ for $U_2$ and $U_1$, respectively. Specifically, the DO of $U_2$ is $2$, while $U_1$'s is $4$, $6$, and $8$ when the number of antennas is $N=1$, $N=2$, and $N=3$, respectively.

\begin{figure} [h]
	\centering
	\includegraphics[width=2.5in]{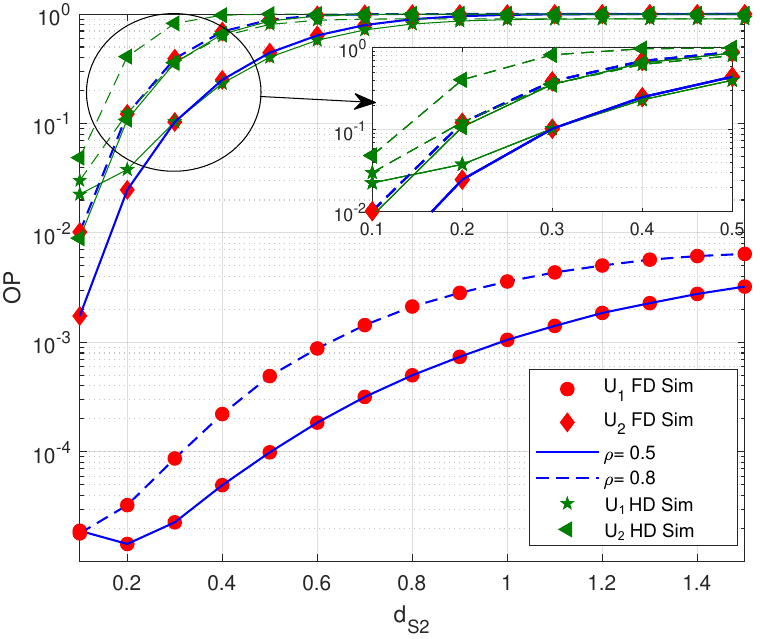}
	\caption{The OP performance versus $d_{S2}$ of the system.}
	\label{fig4}
\end{figure}

Fig. \ref{fig4} illustrates the impact of the distance between the BS and the near user on the OP performance of users in the investigated system with direct link. 
Curves are presented for different values of $\rho$ and $P_S=-20$ dB. 
As observed, for HD and FD as $d_{S2}$ and $\rho$ increases, OP performance of both users degrades. 
Due to the limited space, we cannot present the figure of the OP performance versus $\rho$. When $\rho=0$, no energy is harvested. 
For $0<\rho<1$, as $\rho$ increases, the harvested energy increases and the outage performance decreases for both users.

\section{Conclusion}
This study investigates the OP of an Alamouti/MRC structure in a single-phase downlink multi-user MIMO-NOMA system. 
In the analysis of the proposed system, SINR expressions have been obtained, and closed-form expressions for the OP have been derived. 
Overall, the far user performs considerably, especially in the presence of direct link and FD mode.

\section*{Acknowledgments}
This study was supported by Scientific and Technological Research Council of Turkey (TUBITAK) under the Grant Number 121E467. The authors thank to TUBITAK for their supports.


 




\bibliographystyle{IEEEtran}
\bibliography{IEEEabrv,References}
\vfill

\end{document}